# Unraveling ultrashort laser excitation of nickel at 800 nm wavelength

T. Genieys[1], M. N. Petrakakis[2,3], G. D. Tsibidis[2,4], M. Sentis[1], O. Utéza[1]

[1] Aix-Marseille Univ, CNRS, LP3 UMR 7341, F-13288 Marseille, France
[2] Institute of Electronic Structure and Laser (IESL), Foundation for Research and Technology (FORTH)
N. Plastira 100, Vassilika Vouton, 70013, Heraklion, Crete, Greece
[3] Materials Science and Technology Department, University of Crete, 71003 Heraklion, Greece
[4] Physics Department, University of Crete, 71003 Heraklion, Greece

E-mail: genieys@lp3.univ-mrs.fr

## Abstract

The optical response of nickel is studied in a wide range of laser fluence, below and above the ablation threshold, by self-reflectivity measurements of ultrashort 800-nm single laser pulses. At the ablation threshold, the reflectivity remains unchanged with respect to its unperturbed value irrespective of the pulse duration, from 15 to 100 fs, consistently with the steadiness of the laser-induced ablation threshold fluence $F_{th}$ for all pulse durations tested. Until the ablation threshold ($F \leq F_{th}$) and whatever the pulse duration, the disturbances caused to the initial structure of the electron gas distribution by the laser energy deposition are limited, having no significant impact on the transient optical response of nickel and on its ablation threshold. At higher laser fluences ($F > F_{th}$), the reflectivity becomes rapidly dominated by the contribution to the optical response of the fast-thermalized free electrons (4s-band) with energy largely above the Fermi energy level. In these conditions, the reflectivity decreases for all pulse durations enhancing laser energy coupling and larger optical absorption at the surface of nickel. The optical response of nickel under ultrashort (15-100 fs) irradiation is thus fully elucidated on a wide range of fluence (0.3 $F_{th}$ - 30 $F_{th}$) and for pulse duration down to few-optical-cycle pulse duration. As a key parameter for benchmarking laser-matter interaction in poorly known conditions yet, the evolution of the effective electron collision rate is determined as a function of fluence and pulse duration in very good consistency with experiments.

## 1. Introduction

Optical response of materials exposed to ultrashort laser visible or near-infrared pulses closely depends on the evolution of the electronic configuration of outer layers of the electron distribution of states [1]. In metals, the general picture is that conduction free electrons gain energy from photons of the laser field through Inverse Bremsstrahlung absorption mechanism. The electron energy is further redistributed in the electron subsystem through electron-electron collisions on typically tens or hundreds femtosecond time scale [2] and progressively transferred to the surrounding lattice through electron – phonon coupling on a picosecond time scale [2-4]. For femtosecond pulses, these energy exchange mechanisms are largely separated in time. To account for such non-equilibrium situation, the two-temperature model (TTM) [5] which simulates the temporal and spatial temperature evolution of free electrons and ion subsystems is widely employed to describe the laser interaction with metals. This model assumes an instantaneous thermalization of the electronic distribution to a single temperature with energy redistribution to the electron and lattice subsystems governed by two differential equations linked by a coupling parameter related to the electron-phonon scattering rate. However, in many situations, this approach has been proven questionable, especially when studying ultrashort laser excitation of metals for which the time for electronic distribution thermalization can require a few hundred femtoseconds [2,6-8]. In general, the assumption of an instantaneous thermalization leads to an overestimation of the electronic temperature and an inaccurate estimation of the optical transient properties and of the dynamics and magnitude of the energy transfer from the electron subsystem to the lattice [2,7-9]. Improvement of the theoretical description involves understanding of complex phenomena at different time scales, from femtoseconds to nanoseconds. It comprises laser excitation at the pulse (femtosecond) time scale and further energy transfer and dissipation in the material at longer (picosecond and nanosecond) time scales, eventually yielding to its macroscopic transformation (melting, ablation) when the energy deposition is intense enough [10-12]. The capability to efficiently describe the material response to ultrafast processes through a detailed thereotical model is expected to allow control of the energy deposition which is a very important aspect in laser processing [13].

However, the interaction of ultrashort laser pulses (down to few-optical-cycle pulse duration) with metals is still unexplored. Indeed, a preliminary obstacle is to manage the propagation of such ultrashort and large spectral bandwidth laser pulses to the area of interest. Recently, experimental works addressed the determination of laser damage threshold fluence of optical materials and components including metallic mirrors in view of robust operation of high peak power laser installations [14]. On a wider perspective, experimental developments were intended to demonstrate ultrafast switching capability like ultrafast (de)-magnetization in metals [15,16] or electronic signal processing in solids at extremely (PHz) high speed [17]. Considering the topic of ultrafast laser heating of metals, important insights using pump and probe approach point out the influence of the electronic structure on thermalization of electron subsystem and energy exchanges with lattice [6,7,16,18-20]. For instance, in noble metals like silver or gold, it was shown the importance of screening ascribed to the d-band electrons and their importance in the thermalization of the free electron gas [7,19,21]. Many theoretical and numerical developments enriched the experimental findings with detailed contributions to the understanding of the evolution of the thermophysical parameters of the irradiated metals and of their optical and thermal response in highly non-equilibrium conditions [5,8,11,22-27]. In the particular case of nickel, it was shown using time-dependent density functional theory the formation of two subsystems of electrons. Each one was achieving its own temperature which results in delaying the thermalization time of the whole electronic system [28]. Importance of the d-band of its electronic structure was also postulated in the process of energy exchange and redistribution with consequences on the building of its transient optical response and of its final physical state [24,28-31]. As indicative examples, an electron thermalization time of 80 fs was measured on nickel films irradiated with 85 fs Ti:Sa pulses [30]. It was also shown that strain generation and lattice thermalization were sensitive to the excitation of the d-band bound electrons when performing simulations with excitation pulses in the 10 fs – 2 ps range [29]. In [32], it was demonstrated that the electron energy band structure evolves with the fluence of the femtosecond pulse, which strongly impacts its dielectric function and further the formation and characteristics of ripples.

In this work, we focus on the electronic heating of nickel by an ultrafast laser with particular attention to its optical response at the pulse time scale. To progress in the understanding of energy coupling in nickel in ultrashort regime, we conduct self-reflectivity measurement below and above the ablation threshold (section 2), which is an appropriate tool to study the response of metals to ultrashort laser excitation [19,33-35]. The range of pulse duration tested (15 – 100 fs) is attractive to study because of its closeness to the characteristics of the electron – electron population thermalization [30] with potential effects on the energy coupling to the lattice and on the transient optical properties. The experiments thus give access to observables depending on the incident fluence and pulse duration. They provide pertinent information (section 3) to unravel the optical response of nickel in the femtosecond-laser ablation regime as a function of laser excitation and to determine the evolution of the effective electron collision rate (incorporating electron-electron and electron-phonon contributions) as an important benchmarker of the interaction.

## 2. Experimental results

### 2.1 Laser-Induced ablation threshold fluence in ultrashort regime

All irradiation experiments are performed in air at normal incidence and in single-shot regime. The beam line delivers linearly polarized 1 mJ - 30 fs pulses at 800 nm (1.55 eV) central wavelength ($\Delta\lambda \cong 760 - 840$ nm FWHM). To get longer pulse durations, the beam is pre-chirped through compressor grating adjustments. For the 30, 50 and 100 fs cases, the beam is focused onto the nickel target by a metal-coated 90° off-axis parabola of 152.4 mm focal length. The beam waist at $1/e^2$ is measured by beam imaging in the focal plane where the target surface is located ($\omega_0 = 11$ µm) (see Figure 1). To access few-optical-cycle pulse duration (15-fs case), cross-polarized wave technique is implemented [36] to broaden the beam spectrum ($\Delta\lambda \cong 720 - 880$ nm). Pulse compression to ultrashort pulse duration is achieved using chirped mirrors and a pair of fused silica wedges. To get similar beam dimensions on target, a shorter metal-coated 90° off-axis parabola of 52.8 mm focal length was used ($\omega_0 = 7.75$ µm, see Figure 1) [37].

The laser-induced ablation threshold fluence ($F_{th}$) of nickel samples (GoodFellow Ni000624, high purity 99.99+%, thickness of 3.2 mm) is determined for pulse duration of 15, 30, 50 and 100 fs. Considering Gaussian pulse spatial distribution, the usual diameter-regression technique [38] is used to infer the threshold energy $E_{th}$ (see Figure 1). The corresponding peak threshold fluence is then calculated from the formula: $F_{th} = \frac{2E_{th}}{\pi \omega_0^2}$. The values of the ablation threshold fluence are shown in table 1 for each pulse duration. The ablation threshold fluence of nickel does not vary for pulse duration changing from 15 to 100 fs similarly to what we observed on other ultrashort laser-irradiated metals [37,39]. It is equal to $F_{th} \cong 0.33$ J/cm² which is in good agreement with previous reports performed in similar operating conditions (Ti:Sapphire laser @ 800 nm in air). For instance, using 90-fs single-shot pulses, K. Zhang et al. [40] determine on high purity mm thick Ni samples: $F_{th} = 0.36 \pm 0.02$ J/cm², which closely matches our experimental determination. Other experiments yield: $F_{th} = 0.384 \pm 0.03$ J/cm² at 150 fs [41], and $F_{th} = 0.41$ J/cm² at 100 fs [42]. In this work, the discrepancy with our result may be attributed to the use of high numerical aperture focusing making difficult the precise evaluation of the beam waist size.

Note that measuring a constant ablation threshold in the whole range of pulse duration explored (15 – 100 fs) prolongs similar observations made on several metals excited by femtosecond pulses [4,43,44] to the few-optical-cycle regime.

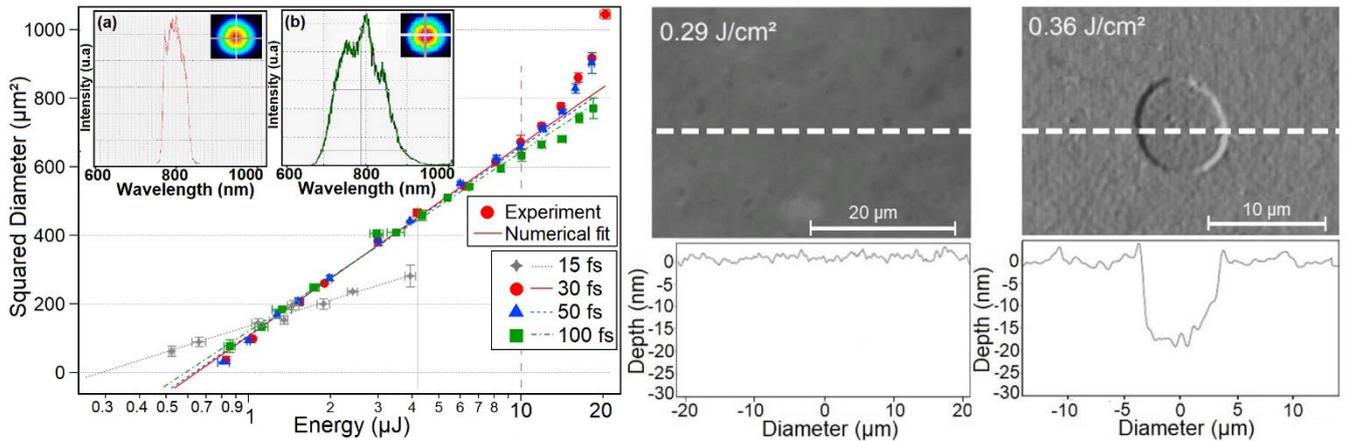

Figure 1. (Left) Evolution of squared diameter D² versus incident laser energy for each pulse duration. The vertical error bars represent the standard deviation over 8 measurements. The horizontal error bars correspond to shot-to-shot incident energy fluctuations measured on a photodiode. The grey-solid and red-dash vertical lines mark the onset of nonlinear effects in air at 15 and 30 fs as measured beforehand [37]. The insert shows the spectral and beam waist spatial characterization for the 30-100 fs (a) and 15-fs cases (b) respectively. (Right) 2D and 1D profiles captured by confocal optical microscopy below and above the ablation threshold fluence. The data are given at 15 fs. Similar data have been obtained for the other pulse durations.

| Pulse Duration | 15 fs | 30 fs | 50 fs | 100 fs |
|---|---|---|---|---|
| $F_{th}$ (J/cm²) | 0.33 ± 0.03 | 0.33 ± 0.02 | 0.33 ± 0.02 | 0.32 ± 0.02 |

Table 1. Ablation threshold fluence of nickel for 15, 30, 50 and 100 fs. The uncertainty in $F_{th}$ determination is $\cong \pm 0.03$ J/cm² for 15 fs pulses and $\cong \pm 0.02$ J/cm² for 30 – 100 pulses, accounting for laser energy fluctuations and the precision error of the measurement of the beam waist $\omega_0$. The error bar is slightly higher for the 15-fs case due to higher laser energy fluctuations measured in this laser operating condition.

*2.2 Measurement of reflectivity in ultrashort regime*

The evolution of the reflectivity of the target has been measured under diverse conditions of ultrashort laser excitation. For each single shot at a given energy (fluence), the incident and reflected energies of the pulse have been measured by photodiodes allowing the determination of the reflectivity R integrated over the pulse duration. The evolution of reflectivity was further normalized by setting R = $R_{0,mes}$ at very low incident fluence ("unperturbed" material state) for which no variation of Fresnel reflection coefficient is detected. To determine this quantity $R_{0,mes}$, the collimated beam line operated at 30 fs was set at very low incident energy. In those irradiation conditions no changes in temperature (electronic and lattice) and permittivity (thus conductivity) occur. The measured value is: $R_{0,mes}$ = 0.686 +/- 0.012. The measurement uncertainty is due to the laser energy fluctuations, scattering and precision of the response of the measuring device. The quantity $R_{0,mes}$ is in good agreement with other experimental measurements ($R_{0,exp}$ = 0.689, [45]) and calculations using the dielectric function coefficients of nickel at 800 nm ($R_{0,calc}$= 0.685 [46]).

The measurements of reflectivity as a function of incident fluence and for the four pulse durations studied are shown in Figure 2. For all cases, the reflectivity remains constant (equal to its unperturbed value) until a given fluence $F_{\Delta R} \sim 0.5$ J/cm² for 15 fs, ~ 0.81 J/cm² for 30 fs, ~ 1.04 J/cm² for 50 fs and ~ 1.58 J/cm² for 100 fs, with the criterion applied: $\Delta R = R_{0,mes} - R > 0.012$ (marked by colored arrows in Figure 2). For fluences higher than $F_{\Delta R}$, the reflectivity starts to decrease. When the pulse duration is shortened, for a fixed given normalized fluence, the reflectivity decreases to lower values. The fluence $F_{\Delta R}$ is also weakly decreasing with the shortening of pulse duration ($F_{\Delta R,15fs} \sim 0.3\ F_{\Delta R,100fs}$). When a larger intensity is applied, higher free electron temperature is expected [47,48] resulting in variation of the transient optical properties and reduction of reflectivity.

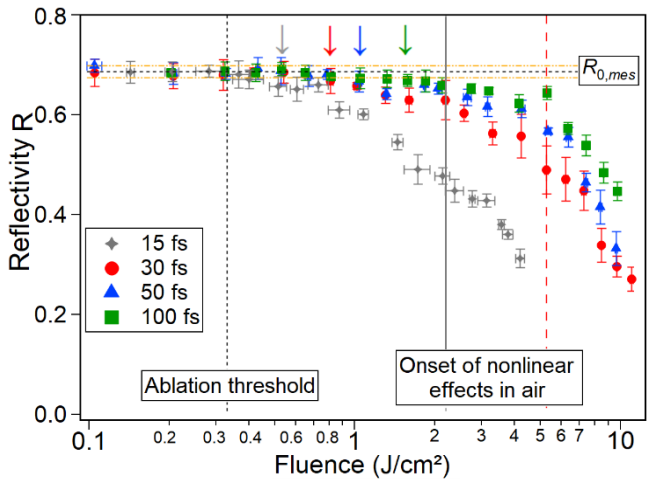

Figure 2. Evolution of reflectivity R of the nickel sample integrated over the pulse as a function of the incident fluence for pulse durations from 15 to 100 fs. The horizontal line corresponds to the reflectivity of the unperturbed sample ($R_{0,mes} = 0.686$; the two dotted yellow lines indicate the amplitude of the measurement uncertainty). The colored arrows indicate the fluence $F_{\Delta R}$ from which the reflectivity starts to decrease for every pulse duration.

At the ablation threshold fluence, no change of reflectivity R is measured with respect to its unperturbed value for all pulse durations tested. Despite high intensity reached with ultrashort pulses (> $2\times10^{13}$ W/cm² at threshold for 15 fs), it is assumed that no change in plasma frequency occurs in this low excitation regime [49]. It is also supposed that the electronic temperature reached for the different pulse duration does not induce detectable modification of the optical properties of the material during the pulse. This will be justified later in section 3.3. The absorbed fluence is thus the same for the different pulse durations which supports the determination of a constant ablation threshold fluence. Similar results have been observed for other transition and post-transition metals including copper, tungsten and aluminum [37,39].

## 3. Optical response of nickel under femtosecond irradiation

### 3.1 Phenomenology of ultrafast laser optical coupling in Ni: role of 3d-band bound electrons

Nickel is a transition metal having a peculiar electronic configuration ([Ar] $3d^8 4s^2$). Its electronic d-band is almost full, leading to a high density of electrons up to the Fermi energy (Figure 3a). Upon laser excitation, this implies a subtle interplay between interband and intraband transitions [23,24,49] as conceptually schematized in Figure 3.

Firstly, we describe the phenomenology of ultrafast laser heating of nickel corresponding to experimental conditions met at low F < $F_{th}$, and moderate F ≅ $F_{th}$ excitations or slightly above $F_{th}$ (and also corresponding to the beginning of high excitation F > $F_{th}$), for which we did not detect any significant modifications of the reflectivity compared to the unperturbed state (Figure 3b). In the initial steps of laser heating, photons couple with 3d-band electrons close to the Fermi energy to promote them at higher energies in the 4s-band where there are many vacant states (see the transitions associated to the two green arrows in Figure 3b). Using data from Figure 2 and optical characteristics of our experiment, we determine the energy absorbed $Q_{abs}$ ($Q_{abs} = (1-R) \times E_{inc}$) and the corresponding absorption volume V to estimate the number of photons absorbed in average per electron $n_{ph/el}$ in the focal volume. The absorption volume is defined as a cylinder of radius $\omega_0$ and of axial extension corresponding to the optical penetration depth $l_{pen} = \frac{c}{2\omega\kappa} = 14.6$ nm with $\kappa = Im(\sqrt{\varepsilon}) = 4.37$ for nickel at 800 nm [45], ($\varepsilon$ standing for the dielectric function). For nickel, the number of conduction electron participating to laser coupling is progressively increasing due to the possibility of mobilizing the reservoir electrons located in the 3d-band [49]. For experimental conditions corresponding to the figure 3b, below and close to the ablation threshold, these changes are nonetheless moderate and we reasonably assume $n_e \cong 2 \times n_{at}$ [49,50] (with atomic density $n_{at} = 9.13 \times 10^{22}$ cm⁻³). Considering all these data, the number $n_{ph/el}$ is calculated below unity, equal to 0.78 photon per electron. Following one-photon Inverse Bremsstrahlung absorption, it thus promotes the formation of an excited electron population around the Fermi energy (grey hatched zones in figures 3b,c). Since the kinetic energy communicated to the electron is lower than the photon energy ($k_B T_e < h\nu$), this electron population extends in the energy region $E_F - h\nu < E < E_F + h\nu$, where E and $E_F$ indicates the electron energy and the Fermi energy, respectively. Importantly in nickel, the 3d-band electrons are much heavier than the free electron population, with an effective mass being ~ 6 times larger than the free electron mass [51]. So, despite increase of their kinetic energy through photon coupling (however modest as estimated by the calculation of $n_{ph/el}$), they acquire little additional mobility. As a result, they weakly interact with the other electrons. Thereby, this electron population has been labelled "nonthermal" [2,8,23,28,29], and in this range of low to moderate excitation (F ≅ $F_{th}$), the thermalization time dominated by the contribution of this nonthermal population is long [2,23,28]. According to the Fermi liquid theory, simplified calculations done in the framework of random phase approximation on the nonthermal population [23] ($\tau_{ee} \cong \tau_{e0}(\frac{E_F}{E-E_F})^2$ (eq. 1), where $\tau_{e0} = \frac{128}{\sqrt{3}\pi^2 \omega_p}$ (= 0.31 fs), and $\omega_p$ = 2.425×10$^{16}$ s$^{-1}$ is the plasma frequency) yields an electron-electron collision time averaged over the accessed electron energy range of ~ 50 fs.

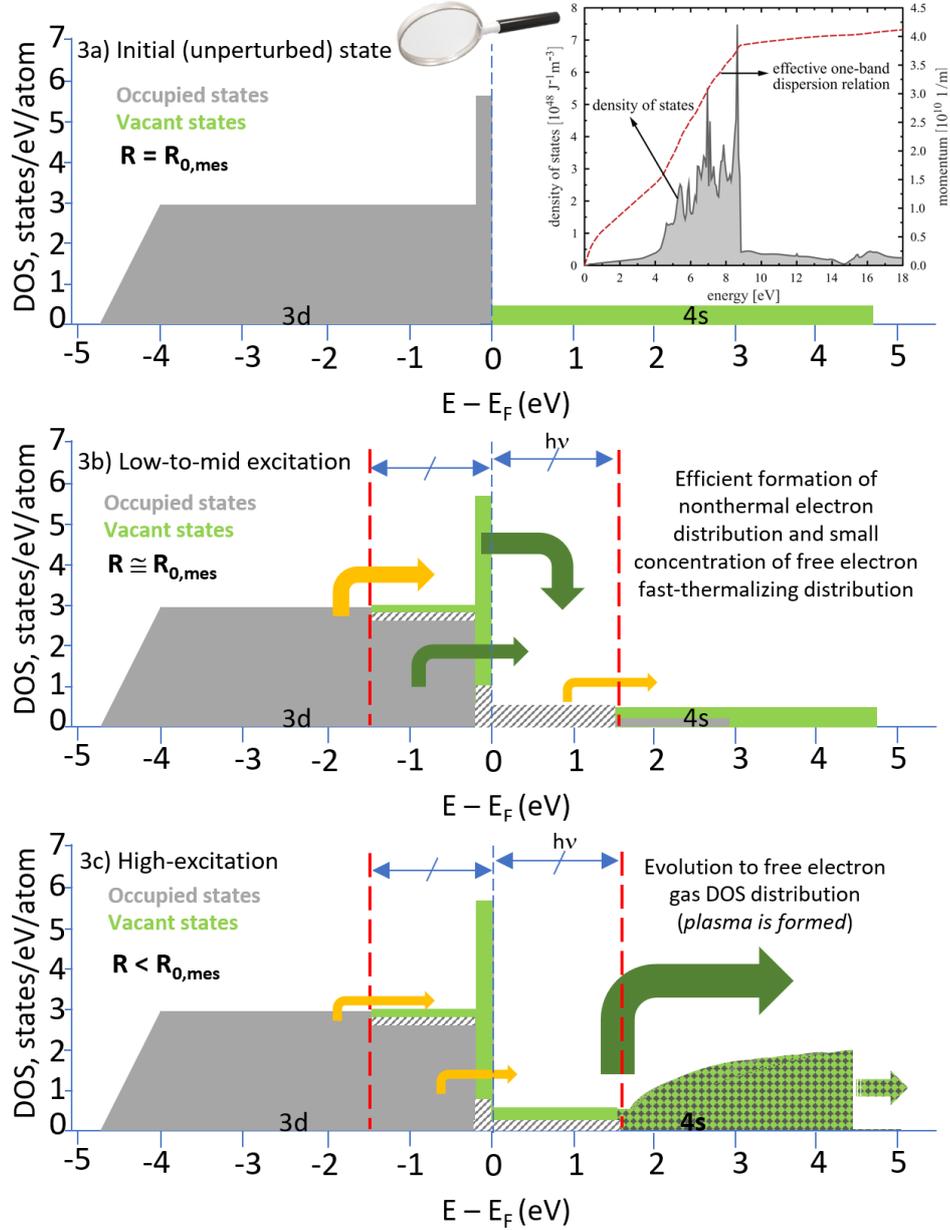

Figure 3. Schematic representation of the evolution of the nickel density of states (DOS) at different excitation levels. The insert in figure 3a (initial unperturbed situation) shows the calculated DOS of nickel and the corresponding averaged dispersion relation in the framework of the effective one-band model (taken from [8]). In figures 3b,c (excited nickel situations), the colour and thickness of the arrows conceptually refer to the effectiveness of the electron-electron transition because of easiness of momentum conservation and of Pauli exclusion principle: "high effective" transition in green, "low effective" transition in orange; the difference in width of the arrow is in accordance with the number of transitions that can occur according to the volume of the phase space available for the electron-electron scattering events (see text). The grey hatched zones indicate the formation of the nonthermal electrons. In figure 3c, the green and grey dotted region corresponds to the transition of the Ni DOS to free electron gas plasma distribution at high excitation with high increase of the volume phase space available for the free electron population.

Considering that a number of collisions between charge carriers, typically of the order of ten [52], is required to transfer the energy and to reach thermal equilibrium of the excited electron distribution, this yields a thermalization time of ~ 500 fs, in agreement with experimental estimates (~ 260 fs) made in weak excitation conditions ($\cong$ 10 mJ/cm² << $F_{th}$) on thin nickel films [53]. So, the thermalization time appears to be much superior to all pulse durations used here. This means that due to weak interaction between nonthermal electrons no complete thermalization of the whole electron population is achieved during the pulse at and close to the ablation threshold. Note that the Fermi energy ($E_F$ = 11.7 eV) was calculated from [54] considering two conduction electrons per atom: $E_F = \frac{m_e v_F^2}{2}$, with the Fermi velocity $v_F = \hbar k_F/m_e$ and the Fermi wave vector $k_F = (3\pi^2 n_e)^{1/3}$; and the accessed energy phase space was taken flat and covering the full one-photon absorption range for simplification ($E_F$ – 1.55 eV < E < $E_F$ + 1.55 eV) and without any consideration of momentum space (all transitions permitted).

As a result of the formation of the nonthermal electron population, the right part (upper energy edge) of the 3d-band is depopulated, as schematized by the green region below the Fermi energy in Figure 3b. If there is a sufficient number of photons still to be coupled, these vacant states of the 3d-band (below Fermi energy at the right of the 3d-band) could be (re)-populated upon photon absorption by intraband transitions originating from 3d-band electrons of lower energy (Figure 3b, mid-thickness orange arrow). Meanwhile the 3d-electrons initially accelerated to higher energy may also sequentially absorb a second photon upon Inverse Bremmstrahlung absorption and reach higher energy levels in the 4s-band (Figure 3b, small-thickness orange arrow). However, these two electron transitions are little efficient in comparison to those yielding to the formation of the nonthermal electron population because of the low number of photon remaining and of the small phase volume accessible to them. This is conceptually marked by the orange colour of the arrows in Figure 3b, with respectively a low number of arrival states defining the mid-thickness orange arrow (at left in Figure 3b), and a low number of departure and arrival states inducing the small-thickness orange arrow (at right in Figure 3b). Finally, as apparent from Figure 3b and our simplified estimations, the disturbances induced to the structure of the electron gas distribution remain limited in this excitation regime (F $\cong$ $F_{th}$), in agreement with measurements showing the absence of changes of material reflectivity.

When the laser excitation is increasing (F >> $F_{th}$, Figure 3c), the probability of intraband absorption of photons by electrons (from low-to-higher 4s states, green arrow) becomes high due to facilitated accessibility and large availability of electron energy states in the high energy phase space. In those conditions, the chemical potential shifts towards higher energy (Fermi energy smearing) [49], the number of free electrons increases, and the excited medium resembles progressively to a free electron gas plasma (Figure 3c). Likely, the contribution of the nonthermal electron population to optical coupling and material response rapidly becomes weak and even vanishes during the pulse (as schematized by the thin orange arrows in Figure 3c). To estimate electron-electron interaction characteristics, the energy-dependent collision rate (time) of an excited electron with energy E within an electron bath supposed to be thermalized at temperature $T_e$ is determined from the Fermi-liquid theory [8,55]:

$$\frac{1}{\tau_{ee}(E,T_e)} = \frac{\pi^2\sqrt{3}\omega_p}{128 E_F^2} \times \frac{(\pi k_B T_e)^2 + (E-E_F)^2}{1+\exp(-\frac{E-E_F}{k_B T_e})} \quad \text{(eq. 2)}.$$

To account for highly non-equilibrium conditions accessed with ultrashort pulses, the electronic $T_e$ and lattice $T_i$ temperature evolutions during the pulse (*t* coordinate) and at the beam center are modelled in the framework of the two-temperature approach in which instantaneous thermalization of the free electron gas distribution is assumed [5]:

$$\begin{aligned} C_e \frac{\partial T_e(x,t)}{\partial t} &= -g\,(T_e(x,t) - T_i(x,t)) + S \\ C_i \frac{\partial T_i(x,t)}{\partial t} &= g\,(T_e(x,t) - T_i(x,t)) \end{aligned} \quad \text{(eqs. 3)}.$$

The electron and lattice diffusion terms taking place on picosecond time scale have been omitted because we are focusing our analysis to short time scale (pulse duration) only where effects related to diffusion processes are negligible. $C_e$ and $C_i$ stand for the heat capacities of the electrons and the lattice, *g* is the electron-phonon coupling factor, and *x* is the spatial coordinate along the optical axis (with *x*=0 corresponding to the surface coordinate). The variation of the parameter *g* (in W/m³K) is taken into account using the electron temperature dependent function determined in the work of Lin et al. [22]. $C_i$ is taken constant because the raise of lattice temperature is extremely modest during the pulse ($C_i = C_{i,300K}$ = 3.945×10⁶ J/m³K) [56]. At the ablation threshold, we indeed calculate an increase of $\Delta T_L$ ~ 94 K at the end of the pulse for the (maximizing) 100 fs case. The quantity $C_e$ is recalculated considering its evolution as a function of the electronic temperature defined in [22]. Finally, the source term *S* refers to the amount of laser energy locally absorbed $S(x,t) = \alpha(1-R)I(t)\exp(-x\alpha)$, where $\alpha$ is the inverse of the optical penetration depth, R the reflectivity, and *I(t)* the incident laser intensity for which Gaussian temporal shape is considered. As a first approximation, the absorption coefficient $\alpha$ and the reflectivity R are taken constant. This assumption is fully valid for fluences below, equal and slightly above to the ablation threshold fluence because the reflectivity does not vary significantly (R $\cong$ $R_{0,mes}$ for F $\leq$ 2×$F_{th}$) for all the pulse durations as it is experimentally demonstrated in Figure 2. Note that it is not rigorously the case for 15-fs pulse duration for which the reflectivity measured is slightly below $R_{0,mes}$ ($\Delta$R ~ 5%), this value being entered in the calculations related to Figure 4.

Following this approach, the evolution of the electron-electron collision time as a function of the electron energy at the incident fluence of 2×$F_{th}$ for all the pulse durations and at two different depths (surface and edge of the skin depth) is illustrated in Figure 4.

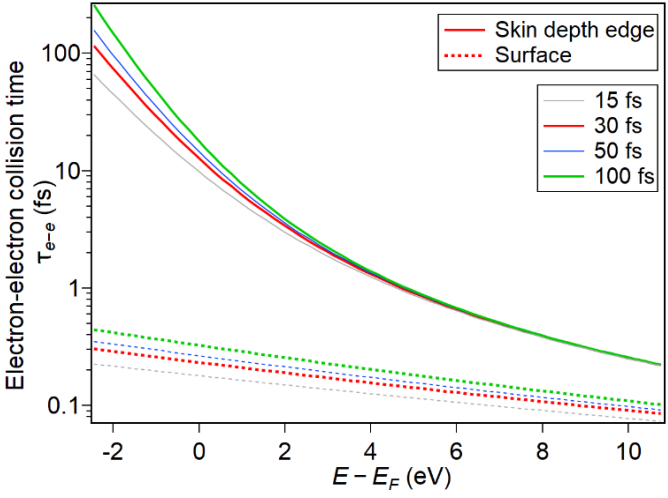

Figure 4. Evolution of electron-electron collision time (in fs) as a function of excited electron energy (expressed with respect to Fermi energy) and for different electron temperature of thermalization of the free electron bath corresponding to experimental conditions encountered at 2×$F_{th}$ at two depths (surface and inner edge of skin depth $x = l_s$) for the four pulse durations studied.

It shows a strong decrease of the electron-electron collision time with the increase of electron energy and it immediately appears that the thermalization time of the electron distribution will be considerably reduced compared to weak excitation (F ≤ $F_{th}$) for which the contribution of the nonthermal population is predominant. In particular, at the surface, the thermalization time will be reduced below the pulse duration when laser excitation is growing due to the swift increase of the electron temperature [8] and the progressive importance of the free electron population.

## *3.2 Dielectric permittivity as a tool to assess the transient optical response of nickel*

The changes in the electron state occupation due to laser excitation are expected to have an impact on the optical response of the material. In our experimental approach, the optical parameter values are monitored by means of the measurement of the pump pulse signal reflected by the sample. The evolution of the electron state occupation also leads to a modification of the dielectric permittivity of the material. Assessing the dielectric permittivity thus serves as a tool for (re)-calculating the reflectivity of the excited system to compare with the experiments. Here, we assume that all the electronic transitions on 3d- and 4s-bands are permitted. Moreover, we only consider linear absorption, corresponding to one-photon Inverse Bremsstrahlung absorption. We argue these assumptions as follow. Firstly, the changes induced by the optical coupling in our excitation conditions primarily concern the outer electron states of the distribution of states for which the transitions are allowed and momentum conservation is easily fulfilled. Secondly, the number of photons absorbed in average per electron was calculated equal to 0.78 photons per electron at the ablation threshold. So, the hypothesis of keeping the absorption linear holds even in ultrashort irradiation regime and rather high incident intensity reached in this case (> 2×10$^{13}$ W/cm² for 15 fs pulse at threshold). Note that this calculation also supports the creation of a large number of excited electrons close to the Fermi energy (nonthermal electrons) at moderate excitation in agreement with the phenomenology described in section 3.1.

To account for the different channels of laser coupling in nickel and in particular for the contribution to the reflectivity of the interband and intraband transitions, we consider the Drude-Lorentz model. The dielectric permittivity is expressed by the following equation [46,57]:

$$\varepsilon_{Ni} = \varepsilon_r + i\varepsilon_i = \varepsilon_{intra} + \varepsilon_{inter}$$
$$= [1 - \frac{f_0 \omega_p^2}{\omega(\omega - i\nu)}]_D + [\sum_{j=1}^{k} \frac{f_j \Omega_p^2}{(\omega_j^2 - \omega^2) + i\omega\Gamma_j}]_L \quad \text{(eq. 4)}.$$

The first term $\varepsilon_{intra}$ is conveniently described by the Drude model with $\omega_p$ the plasma frequency associated with intraband transitions with the oscillator strength $f_0$, and the collision rate $\nu$, and $\omega$ is the laser frequency. The second term $\varepsilon_{inter}$ is described by the Lorentz component where $f_j$, $\Omega_p$, $\omega_j$, and $\Gamma_j$ are, respectively, the oscillator strength, the plasma frequency, the frequency and the scattering rate of the harmonically bound electrons that are excited via interband transition j. All the parameters entering the calculations for nickel at room temperature have been defined in the experimental works of Rakic et al. [46]. The plasma frequencies are set to $\omega_p = \Omega_p = 2.425 \times 10^{16}$ s$^{-1}$ in which the number of free electrons per atom is assumed to remain equal to two. This is reasonable considering results of density functional theory calculations for nickel. Indeed they show that the number of free electrons varies from 1.4 at $T_e = 0$ K to 3 at $T_e = 10^5$ K [49,58], which correspond to electron temperature conditions reached in our experiments. Afterwards, the reflectivity is calculated using the following formula [59]: $R = \frac{(1-n)^2 + k^2}{(1+n)^2 + k^2}$ (eq. 5), in which the real (n) and imaginary (k) parts of the complex refraction index are classically defined by:

$$n = \sqrt{\frac{\varepsilon_r + \sqrt{\varepsilon_r^2 + \varepsilon_i^2}}{2}} \quad \text{and} \quad k = \sqrt{\frac{-\varepsilon_r + \sqrt{\varepsilon_r^2 + \varepsilon_i^2}}{2}} \quad \text{(eqs. 6)}.$$

We first consider the reflectivity for an unperturbed material. In that case, for the unexcited cold solid at room temperature (300 K), the collision rate is governed by the electron-phonon collision rate which is approximated for temperatures in the range from the Debye temperature up to the melting point by the following expression [60]: $\nu_{e-ph} = \frac{3}{2} C_\omega \frac{k_B T_i}{\hbar}$ (eq. 7). $T_i$ is the lattice temperature, and $C_\omega$ is a dimensionless proportionality coefficient which depends on the material properties and that can be determined from experiments. Using the reflectivity measured beforehand for the unperturbed sample ($R_{0,mes} \cong 0.686$) and the Drude-Lorentz model above for calculating the reflectivity, we determine $\nu_{e-ph} \cong 10^{14}$ s$^{-1}$ and $C_\omega = 1.69$. It corresponds to the complex refractive index: $n_{Ni} = 2.43 + i4.38$ which is in good agreement with tabulated values in [54,61] in the corresponding spectral range ($n_{Ni,1.6eV} = 2.43 + i4.31$).

### *3.3 Reflectivity of nickel at the ablation threshold*

The aforementioned approach and formulas are then used to estimate the change of reflectivity accumulated during the pulse at fluences close to the ablation threshold. The modelling approach is worked out in order to match the geometry and temporal arrangement of the measurement (capture of the time- and space-integrated reflected pulse energy). A Gaussian transverse shape dependence to the effective collision rate (calculated at the centre of the beam) is applied and the results are further integrated along the transverse coordinate of the laser beam. The time history (allowing further integration and calculation of the time-integrated reflectivity) is considered by handling the temporal dependence of all the parameters entering in the TTM calculations (eqs. 3).

As described above for fluences $F \cong F_{th}$, the laser irradiation of nickel leads initially to the creation of a large population of nonthermal electrons close to the Fermi energy level and of a small concentration of thermalized free electrons (4s-band), as expected from the estimation of the number of photon absorbed per electron $n_{ph/el}$ at threshold. Those 4s-band electrons are issued from sequential absorption of two or more photons and they are occupying energies largely above the Fermi energy level. In order to describe the energy exchanges between the electron and ion subsystems and to calculate the Drude-Lorentz permittivity $\varepsilon_{Ni}$, the effective electron collision rate $\nu_{eff}$ is determined. It includes the electron-electron and electron-phonon interactions associated to the two electron population components defined previously. Thereby, the effective collision rate is expressed in the following form:

$$\nu_{eff} = (A_{nth}\nu_{e-ph,nth} + B_{th}\nu_{e-ph,th}) + (A_{nth}\nu_{e-e,nth,eff} + B_{th}\nu_{e-e,th,eff}) \quad \text{(Eq. 8)}$$

Where $\nu_{e-e,nth,eff} = \frac{\nu_{e-e,nth-nth} + \nu_{e-e,nth-th}}{2}$ and $\nu_{e-e,th,eff} = \frac{\nu_{e-e,th-th} + \nu_{e-e,th-nth}}{2}$. In these two expressions, the interaction of each electron population with itself and the other component is described and it accounts for their respective characteristics of thermalization. Note that an equal contribution to the electron-electron collision rate is assumed for simplification as we do not know the details of the electron distribution in energy at any time. The different components entering in the evaluation of the electron-electron collision rate are calculated for every time and space steps. To do that, we use successively the equation 1 for $\nu_{e-e,nth-nth}$ (with nonthermal electron energy range $E_F - 1.55$ eV $< E < E_F + 1.55$ eV) and for $\nu_{e-e,th-nth}$ (with free electron energy range $E_F + 1.55$ eV $< E < E_F + 2 \times 1.55$ eV corresponding to the sequential absorption of two photons per electron), and the equation 2 for $\nu_{e-e,nth-th}$ (with $E_F - 1.55$ eV $< E < E_F + 1.55$ eV and $T_e$ defined by eqs. 3) and for $\nu_{e-e,th-nth}$ (with $E_F + 1.55$ eV $< E < E_F + 2 \times 1.55$ eV and $T_e$ defined by eqs. 3). The electron – phonon collision rate terms ($\nu_{e-ph,nth}$ and $\nu_{e-ph,th}$) are evaluated using the equation 7 and the TTM equations 3. In the equation 8, the percentage coefficients, $A_{nth}$ and $B_{th}$ (with $A_{nth} + B_{th} = 1$), consider the partition between the number of nonthermal and thermalized electrons during laser heating. To account for diverse quantitative population distributions and all possible scenarios, we take variable the coefficients $A_{nth}$ and $B_{th}$. When $A_{nth}$ is equal to 0 or 1 (corresponding to only a single electron population component), the equation 8 is simplified to the sum of their electron-electron and electron-phonon contributions. Importantly, we recall that the evolution of each term entering in equation 8 as a function of progress of laser excitation are considered through the swift increase of the electron temperature and much more modest and progressive increase of the lattice temperature (TTM model, eqs. 3). Due to its nonthermal nature the contribution to the effective collision rate of the nonthermal population slowly evolves during the pulse, contrary to the fast-thermalizing free electron distribution component. Note also that the assumption of quasi-instantaneous thermalization used to establish the TTM model yields a slight (and artificial) increase of the corresponding electron-electron collision rate [7] but without any consequence on our results and interpretation of Figure 5 provided below.

Finally, the reflectivity coefficient (eq. 5) is evaluated at threshold and at the surface ($x=0$ in eqs. 3). It is integrated over the pulse using the Drude-Lorentz model (eq. 4) and the effective collision frequency (eq. 8) which time and space dependence as a function of excitation is accounted for in our modelling approach. In the reflectivity calculation we do not consider any change in the characteristics (oscillator strength, plasma frequency, and scattering rate) of the optical response of the bound (interband) component due to the raise of lattice and electron temperatures. This is reasonable because significant increase of the lattice temperature only occurs much after the pulse termination. Moreover, for fluences around the ablation threshold, the electronic temperature remains lower than the Fermi temperature which corresponds to an excitation regime weakly perturbed (cold solid regime). Obviously, this will become questionable at higher excitation (F > $F_{th}$) when the number of photons coupled to electrons becomes high and the energy deposited in the electron subsystem approaches or exceeds the cohesion limit of the material (see 3.4).

The variation of reflectivity with respect to its unperturbed value is shown on Figure 5 as a function of number partition between the nonthermal and free electron thermalized distributions. It is minimal when a large proportion of the nonthermal electron population is created and persists significantly during the pulse. As an example, $\Delta R << 0.0005$ for 100/0 case and all pulse durations, and corresponding maximum effective collision frequency reached at the end of the pulse amounts to ~ $1.52 \times 10^{14}$ s$^{-1}$ for 100 fs pulse duration. Infinitesimally small differences are observed for the other pulse durations due to the absence of significant raise of the lattice temperature during the pulse (see the insert in Figure 5). Such reflectivity change cannot be detected with our reflectivity measurement system in agreement with observations (Figure 2).

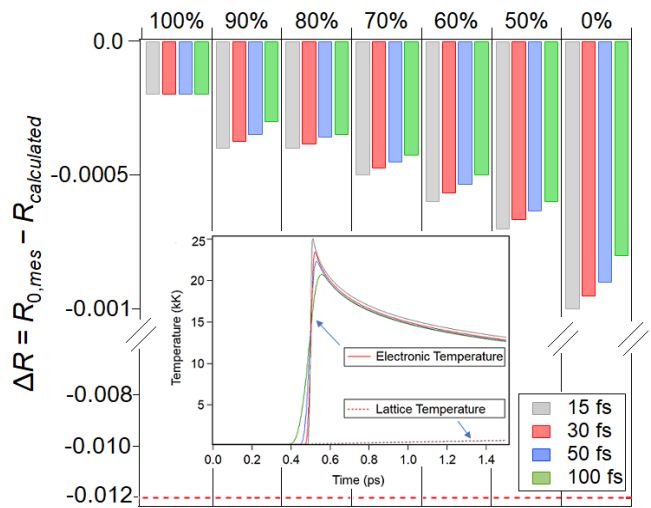

Figure 5. Calculated change of reflectivity (with respect to measured unperturbed value, $R_{0,mes} = 0.686$) at F $\cong$ $F_{th}$ for different number partition between the nonthermal and free electron thermalized distributions (from 100/0 to 50/50 and 0/100). The red dash line indicates the minimum amount of reflectivity change that can be interpreted experimentally as a change induced by the laser excitation. The insert shows the time evolution of the electronic and lattice temperature for all pulse durations calculated using TTM equations 3.

Conversely the variations of reflectivity are maximized when the fast thermalizing free electron population becomes significant and the electron-electron collision rate rapidly dominates the energy exchanges during the interaction. As an example, $\Delta R \cong 0.001$ for 15 fs, 0/100 case, and the corresponding maximum effective collision frequency reached at the peak and centre of the pulse amounts to ~ $9.42 \times 10^{14}$ s$^{-1}$. However, and most importantly, for excitation conditions depicted in Figure 5 (corresponding to conditions at the ablation threshold F $\cong$ $F_{th}$), the change of reflectivity appears to be extremely small in every case. It is thus not surprising that no variation of the reflectivity is experimentally detected at the ablation threshold (and slightly above), as shown in Figure 2.

So, for moderate fluences in the vicinity of the ablation threshold F $\cong$ $F_{th}$, the transient optical response (and related properties of reflectivity and absorption) of nickel is not sensitive to the changes in the electron distribution induced by the laser excitation. In particular, it is not sensitive to the formation of a nonthermal electron population even if the latter promotes a delayed thermalization time of the whole electronic population much longer than the laser pulse duration. This first conclusion holds even for extremely short laser pulse duration ($\leq$ 15 fs) and rather high applied peak intensity (~ $2 \times 10^{13}$ W/cm²).

When considering Figure 5, another outcome is that small differences are observed in the transient optical properties when comparing different pulse durations. As an indicative example, the difference of reflectivity variation between 15 and 100 fs is < 25% for all electron distribution partition scenario depicted in Figure 5 and negligible for the 100/0 number repartition case. As it was also seen experimentally (see Figure 2), those differences found as a function of pulse duration are so tiny that they do not affect the total amount of energy density deposited in the material and also its dynamics of transfer and conversion to internal energy. This observation is not surprising when considering the relatively small maxima of the electron temperature reached for all pulse durations as estimated with the TTM approach at the centre and peak of the pulse: 2.16 eV (15 fs), 2.02 eV (30 fs), 1.925 eV (50 fs), 1.785 eV (100 fs) (see the insert in Figure 5). As a result, the very small variation of the electron temperature over the laser pulse justifies why a constant ablation threshold fluence is determined for the pulse duration range studied, even with few-optical-cycle pulse durations. This observation is also in line with the assumption that the absorption remains linear even when high intensity (> $10^{13}$ W/cm²) yielding to the macroscopic transformation of the material is applied.

### 3.4 Evolution of the effective electron collision rate at high excitation fluences $F_{inc} > F_{th}$

The drop of reflectivity (Figure 2) tends to increase and to be significant (measurable) when the influence of the nonthermal electrons reduces and accordingly when the electronic temperature and the contribution of the free electron intraband population increase (transition to free electron plasma). Interestingly, in the work of Bévillon et al. [24], it is shown that the contribution of the bound d-block (3d-to-4s interband transitions) to the optical properties is prevailing at low temperature but gradually vanishes upon laser excitation because of the strong shift of the chemical potential towards high energy. In this case, the Drude model, describing the response of a free electron gas, could be applied to determine the evolution of the reflectivity. To do so, the electronic temperature is calculated from equation 9 [4], using the reflectivity experimentally measured (Figure 2):

$$T_e(x,t) = \frac{4(1-R_{mes})F_{inc}}{3l_s n_e} \exp\left(-\frac{2x}{l_s}\right) \qquad \text{(eq. 9)}.$$

Where $l_s$ is the varying skin depth ($= 2l_{pen}$) and $n_e$ the varying free electron population density, estimated as a function of excitation in [49]. For high excitations, the influence of the non-thermalized population can be neglected, and only collisions between fast-thermalized electrons are considered to calculate the electron collision frequency. The electronic temperature at the surface of the sample, calculated with Eq. 9, is used to obtain the electron-electron collision frequency from Eq. 2 with an energy integration domain centered at $E - E_F = k_B T_e$. For simplification, the evolution of the lattice temperature during the pulse is neglected. The electron-phonon collision frequency is thus kept equal to the unperturbed value ($\nu_{e\text{-}ph} \cong 10^{14}$ s$^{-1}$). This is not a severe approximation because of the consideration of the pulse time scale only in the present work and because the electron – phonon contribution to the effective electron collision rate is rapidly dominated by the contribution of the electron-electron component in these high excitation conditions. As in section 3.3, the Gaussian transverse shape of the beam profile is applied to determine the effective collision rate (primarily calculated at the centre of the beam). Knowing the evolution of the collision frequency as a function of the fluence, the Drude model (first term of eq. 4) can be applied to obtain the evolution of the reflectivity as a function of the incident fluence for the four pulse durations (Figure 6).

At relatively low excitation, and accordingly small effective electron collision frequency, the reflectivity calculated with the Drude model is higher than the unperturbed value $R_{0,mes}$. As it was studied before, this was expected because it is important to consider the contribution of the bound electrons to reproduce the optical response in these excitation conditions (up to $F_{inc} \cong F_{th}$). However, when the excitation increases above the ablation threshold, the Drude model conveniently describes the optical response of the excited material, based on the free electron (4s) intraband population only. In those excitation conditions, the reflectivity drops below its unperturbed value and the calculation is in good accordance with our experimental results. The agreement between the Drude model and experimental results is observed from lower fluences at shorter pulse durations. The heating of the electronic population being proportional to the intensity, a high electronic temperature is reached for lower fluences at short pulse durations, reducing faster the influence of bound 3d-band electrons.

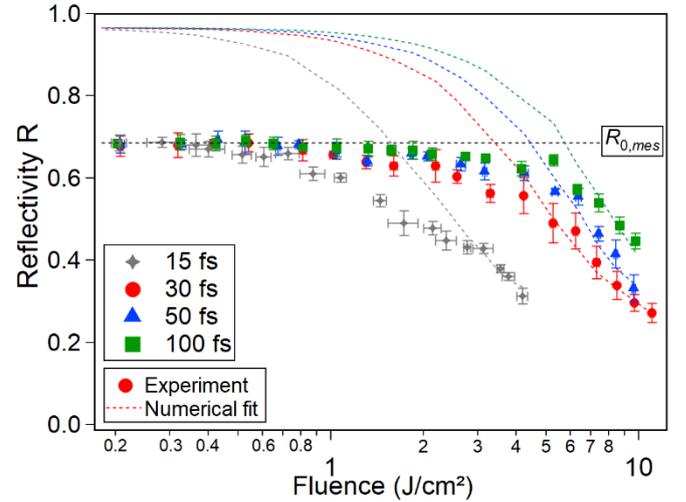

Figure 6. Evolution of the reflectivity calculated with the Drude model, compared to the experimental results for the four pulse durations.

In Figure 7, the calculation of reflectivity using the Drude and Drude-Lorentz models is compared to the experimental results for pulse duration of 30 fs for synthesis purpose. A good initial agreement is obtained using the Drude-Lorentz model. As described in section 3.1, the influence of the band structure remains important until the fluence (and therefore the electronic temperature) increases just above the ablation threshold. However, upon further increase of the excitation (F > 0.8 J/cm² $\cong$ 2.4 $F_{th}$), the disturbances induced to the band structure becomes sufficiently important so that the Drude-Lorentz coefficients are no longer valid. As a result, there is a discrepancy between the calculation and the experiment as shown in Figure 7 for intermediate fluences (0.8 – 3 J/cm² for the 30-fs case). This also qualitatively corresponds to the transient situation depicted in Figure 3c showing the progressive evolution from an optical response first dominated at medium excitation by the bound-electron contribution and further by its excited free-electron component at high fluence. As an indicative marker, it is interesting to approximate the increase of the electronic pressure ($\Delta P_e = n_e k_B T_e$) in the excited material at the fluence of 0.8 J/cm², when the modelling departs from the experiments. Using TTM and the maximum electron temperature reached at the surface and centre of the pulse, the calculation yields to $\Delta P_e \sim 165$ GJ/m³, which exceeds the cohesive energy of nickel (64.9 GJ/m³, [62]). So, it is not surprising that dramatic changes should be incorporated in the description of the Lorentz coefficients entering in the calculation of the dielectric permittivity for improving the modelling approach. This is however out of the scope of the present work.

For the highest excitations (here for fluences ≥ 3 J/cm² at 30 fs), the band structure has no more influence on the evolution of the optical parameters. Indeed, a free electron plasma is formed rapidly during the pulse, and the Drude model incorporating the contribution of the free electron component only must be applied to well reproduce the experiments. Similar observations can be made for the other pulse durations, with the remark that the transition to the free electron plasma state occurs at lower fluences when the pulse duration is shortened due to higher intensity applied (and so higher electronic temperature).

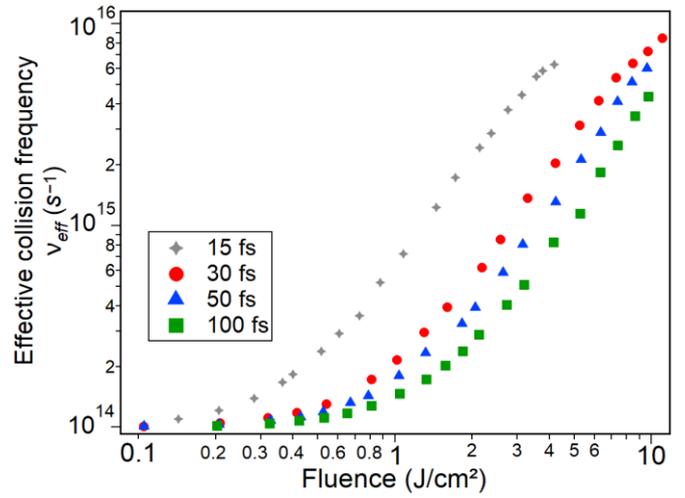

Figure 8. Evolution of the effective collision frequency $\nu_{eff}$ as a function of the incident fluence and for the four pulse durations studied.

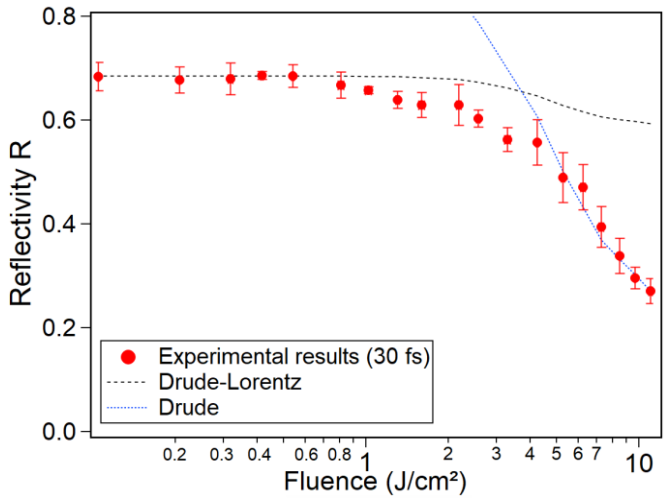

Figure 7. Evolution of the reflectivity vs incident fluence calculated with the Drude model plotted with blue dot line (incorporating the contribution of the free electron component only) and Drude-Lorentz model plotted with black dash line (incorporating the contribution of the bound and free electron components) compared to the 30-fs experimental results.

Finally, as a summary of our works, the Figure 8 consolidates the evolution of the effective electron collision frequency $\nu_{eff}$ as a function of laser fluence and pulse duration. The parameter $\nu_{eff}$ here given is an averaged parameter allowing the accurate prediction of the optical response (dielectric permittivity and further its optical parameters) over an extremely wide range of laser excitation. This is of paramount importance to handle the optical response of a material (here nickel) under femtosecond laser excitation. As an exciting perspective, the development of pump-probe experiment would be interesting to set in to track with high-temporal resolution the transient optical response of nickel upon ultrashort laser excitation. This would allow verifying the quantitative evolution of the effective collision frequency as a function of laser excitation and laser pulse duration, which was closely approached in this work.

## 4. Conclusion

In this paper, a detailed experimental analysis was conducted to describe the optical response of a thick nickel sample irradiated with 800-nm ultrashort single laser pulses. Firstly, the laser-induced ablation threshold fluence was determined in air and in single-shot regime for pulse durations ranging from 15 to 100 fs. It was found constant and equal to 0.33 J/cm² for all pulse durations tested. This is due to the very weak increase of the electron temperature (or in other words to the smallness of the disturbances caused to the electron distribution) reached over the pulse and having no impact on the optical response (reflectivity) of the material whatever the pulse duration. This is verified even when few-optical-cycle pulse durations and high intensity ($\cong 2\times10^{13}$ W/cm²) are applied.

Along with the ablation threshold scaling, we measured the evolution of reflectivity integrated over the pulse duration as a function of the incident fluence. At the fluence ablation threshold and whatever the pulse duration used in the explored range (15 – 100 fs), the reflectivity remains unchanged with respect to the value of reflectivity of the unperturbed sample.

To explain the evolution of the optical response of the material upon increasing laser excitation, we account for the peculiar electronic structure of nickel ascribed to the presence of the bound electron 3d-band extending up to the Fermi energy. This densely populated electron band promotes the formation of nonthermal electrons close to the Fermi energy when the excitation rises, contributing to delayed thermalization of the excited electron gas for fluences up and just above the ablation threshold fluence. However, it also appears that they do not affect significantly the laser energy deposition and the transient optical response of nickel, which is conveniently described by the Drude-Lorentz model in these conditions. In other words, the fact that a nonthermal population is formed and is slowing down the thermalization of the complete excited free electron distribution does not have any noticeable consequence on the strength of laser energy coupling in nickel.

At high excitation regimes ($\gg F_{th}$), the importance of nonthermal electrons is rapidly smeared out and the free intraband (4s-band) population prevails in energy coupling and further in internal energy exchange, transfer and conversion. In particular, much shorter electron thermalization time is reached and the optical response of the nickel sample is conveniently described by the Drude model which quantitatively reproduces the large decrease of reflectivity observed at high excitation.

Finally, as a salient outcome of our works, the optical response of nickel and in particular the evolution of the effective electron collision rate is determined as a function of fluence (0.3 $F_{th}$ - 30 $F_{th}$) and for pulse duration down to few-optical-cycle in very good agreement with experiments. On a modelling perspective, it also appears that simplified classical approaches based on TTM and Drude-Lorentz models correctly predict the evolution of the transient optical properties of the metal (here nickel) at the pulse time scale under ultrashort laser excitation. Those outcomes are of paramount importance to control macroscopic transformation of nickel processed by femtosecond laser pulses.

**Funding**

Financial support of the ASUR platform was provided by the European Community, Ministry of Research and High Education, Region Provence-Alpes-Côte d'Azur, Department of Bouches-du-Rhône, City of Marseille, CNRS, and Aix-Marseille University. M.N.P. and G.D.T. acknowledge financial support from *Nanoscience Foundries and Fine Analysis (NFFA)–Europe* H2020-INFRAIA-2014-2015 (under Grant agreement No 654360), and COST Action *TUMIEE* (supported by COST-European Cooperation in Science and Technology). All the authors thank CNRS-DERCI for their financial support through the International Research Project "*IRP-MINOS*".

**Acknowledgements**

T. Genieys acknowledges the support of *DGA – Direction Générale de l'Armement* (Ministry of Defense), *Aix-Marseille University* and *Ministry of Research and higher Education* for his Ph'D grant.